\documentclass[useAMS,usenatbib,usegraphicx]{mn2e} 

\newcommand{\psrbl}{PSR~B1259-63/LS~2883}
\newcommand{\psrb}{PSR~B1259-63}
\newcommand{\fermi}{\textit{Fermi}-LAT}
\newcommand{\hess}{H.E.S.S.}
\newcommand{\halpha}{H$\alpha$}

\title[Optical spectroscopy of PSR~B1259-63 during 2014]{Optical spectroscopy of PSR~B1259-63/LS~2883 during the 2014 periastron passage with the Southern African Large Telescope\thanks{Based on observations made with the Southern African Large Telescope (SALT) under programs 2013-02-RSA-003 and 2014-01-RSA-001 (PI: B. van Soelen).}}
\author[B. van Soelen et al.]{B. van Soelen$^1$\thanks{Email:vansoelenb@ufs.ac.za}, P. V\"ais\"anen$^{2,3}$, A. Odendaal$^1$, L.~Klindt$^1$, I. Sushch$^{4,5}$, P.J. Meintjes$^1$\\
$^1$Department of Physics, University of the Free State, Bloemfontein, 9300 South Africa\\
$^2$South African Astronomical Observatory, PO Box 9, Observatory 7935, Cape Town, South Africa\\
$^3$Southern African Large Telescope, PO Box 9, Observatory 7935, Cape Town, South Africa\\
$^4$Centre for Space Research, North-West University, 2520, Potchefstroom, South Africa\\
$^5$Astronomical Observatory of Ivan Franko National University of L'viv, vul. Kyryla i Methodia, 8, 79005, L'viv, Ukraine\\}

\begin{document}

\date{Accepted XXX. Received YYY; in original form ZZZ}

\maketitle

\begin{abstract}
The gamma-ray binary system \psrbl\ went through periastron in May 2014.  We report on the optical spectroscopic monitoring of the system from 33~d before to 78~d after periastron, undertaken with the Southern African Large Telescope (SALT).  The H$\alpha$ and He~I ($\lambda$6678) lines exhibit an orbital variation around periastron, with the line strengths reaching a maximum $\sim 13$~d after periastron. The line strength is weaker than observed around the previous periastron in 2010.  There is also a marked change in the line strength and asymmetry around the first disc crossing. These observations are consistent with the disruption of the circumstellar disc around periastron due to the interaction with the pulsar.
\end{abstract}


\section{Introduction}

\psrbl\ is one of only five confirmed gamma-ray binary systems. All of these systems consist of an early main-sequence star with a compact object companion and display non-thermal emission which peaks in the gamma-ray regime \citep[see e.g.][]{dubus13}. Of the five known gamma-ray binary systems, \psrbl\ is unique, since it is the only one where the compact object has been identified as a radio pulsar \citep{johnston92,johnston94}. In the other systems the nature of the compact object remains unclear \citep[for a review of gamma-ray binaries see][]{dubus13}.

Radio observations of the 47.8~ms pulsar allow the binary system's orbital parameters to be well established and the pulsar is in a 1236.72~d ($\sim 3.4$~yr) orbit (with an eccentricity of $e=0.87$) \citep*{wang04,shannon14} around the Be star, LS~2883. \citet{negueruela11} reported on detailed spectral classification of the optical star, estimating a mass of $M_* \approx 31$~M$_\odot$ and a distance of 2.3~kpc to the source.

The system is non-accreting and a shock forms between the pulsar and stellar winds, resulting in particle acceleration and subsequently non-thermal/unpulsed emission \citep{tavani94}.  This emission has been detected from radio to TeV gamma-ray energies and is brightest  around periastron due to the decrease in binary separation \citep[see e.g.][]{johnston05, abdo11, abramowski13,chernyakova14}. Extended radio \citep{moldon13} and X-ray emission \citep{pavlov11,pavlov15} have also been reported.

The position of the circumstellar decretion disc around LS~2883 has been inferred from the eclipse of the radio pulsar between approximately 17 d before until 17 d after periastron \citep{melatos95, johnston96}.  It has been argued that local maxima in the non-thermal (unpulsed) radio  and X-ray emission occur while the pulsar wind is interacting with the circumstellar disc \citep[e.g.][]{chernyakova06,chernyakova14}. Tidal interaction near periastron as well as the effect of the pulsar wind will likely cause the circumstellar disc to be truncated. \citep[e.g.][]{okazaki11}.  Here, we will refer to the phase around the first and last detection of the pulsar as the disc plane crossing.

In 2010, \psrbl\ was, for the first time, observable with the \fermi\ around periastron.  
The source was faintly detected near periastron, but approximately 30 d after periastron the \fermi\ detected a bright gamma-ray flare  \citep{abdo11}, occurring at a phase when the multi-wavelength emission at other wavelengths was already decreasing. This rapid increase was not observed at other wavelengths, including at TeV energies with the \hess\ gamma-ray telescope \citep{abramowski13} and has led to a number of theoretical suggestions \citep*[see e.g.][]{dubus13b,khangulyan12,kong12,takata12,mochol13,sushch14}.

In 2014, \psrb\ went through periastron on 2014 May 4 (MJD 56781.418307) and was observed by a number of different telescopes. Detections have been reported from \fermi, {\it AGILE}, \textit{Swift}/XRT and H.E.S.S.\  \citep*{tam14,malyshev14,tam14b,wood14,pittori14,bordas14,chernyakova15,romoli15}. The \fermi\ flare has also been found to be repetitive, with the flare beginning at the same orbital phase \citep{caliandro15}. However, there was no detection at periastron and the emission was fainter than observed after the 2010 periastron.

The interaction between the pulsar and stellar wind is complicated near the region of the circumstellar deccreation disc that surrounds the Be star. This is expected to introduce variations in the circumstellar disc and subsequently in the associated emission lines originating from the disc. The extent of this variability around periastron has been considered, for example, through smooth particle hydrodynamical simulations by \citet{takata12, okazaki11}, and previously observed by \citet{chernyakova14}. This is also expected (and observed) in other gamma-ray binary systems that contain Be stars: recently \citet{moritani15} reported on monitoring of the TeV gamma-ray binary HESS~J0632+057. Variations in the H$\alpha$ emission line have, for example, also been observed in binary systems such as $\delta$~Scorpii \citep{miroshnichenko01}.

 Here, we report on optical spectroscopic observations undertaken with the Southern African Large Telescope (SALT)  \citep{odonoghue06} located at the South African Astronomical Observatory (SAAO) in Sutherland, showing the variation in the H$\alpha$ and He~I (6678~\AA) emission lines around this period.

\section{Observations}

Spectroscopic observations were undertaken of \psrbl\ using the Robert Stobie Spectrograph (RSS) on SALT \citep{burgh03}. The spectrograph was configured for a wavelength coverage between 6176.6--6983.0~\AA, in order to cover the H$\alpha$ (6656.28~\AA) and He~I (6678.15~\AA) emission lines previously reported \citep{chernyakova14}. The resolution for this configuration was $R = 11020$ at the central wavelength (6613.8~\AA) using a 0.6 arcsec slit. The RSS detector is constructed from three CCDs, which introduces two gaps in the observed spectrum.  The RSS was configured such that both the lines of interest lay on the same CCD for all observations. Each observation consisted of 3 to 4 camera exposures, for a total integration time of between $\sim 476$ and 500~s.

The observation campaign consisted of regular monitoring of \psrbl\ between 2014 April 30 and 2014 July 21 (33~d before to 78~d after periastron respectively) with typically 5--8 d intervals. Observations with more frequent visits were scheduled to occur around periastron and near the time of the reported \fermi\ flare. Unfortunately, poor weather and technical problems prevented the observations around the on-set of the {\it Fermi} flare \citep[e.g.][]{wood14}, though coverage was obtained over most of the periastron period. We present results of 25 observations of the system during this period. 

\begin{figure}
 \includegraphics[scale=.65]{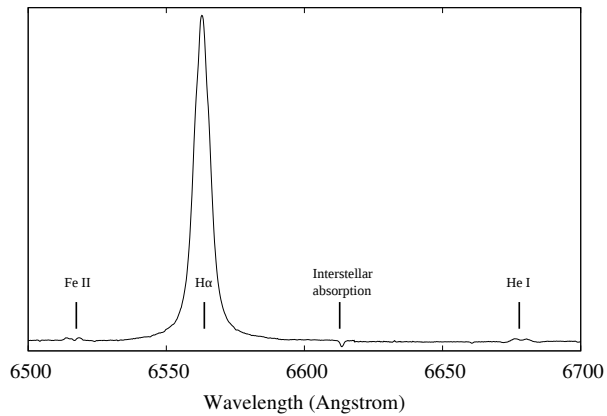}
  \caption{Spectrum of PSR~B1259-63 taken on 2014 March 3.  The H$\alpha$ emission line is clearly present (around 6563~\AA). Also visible are the fainter Fe~II line, He~I line and an interstellar absorption feature.}
 \label{fig:fig1}
\end{figure}

Data reductions followed the standard IRAF procedures in the noao/twodspec package and the exposures for each night were combined to form an averaged spectrum. The spectral flux shape corrections were done using observations of the spectroscopic standard LTT4364 undertaken on 2014 May 11. 
 An example spectrum is shown in Fig.~\ref{fig:fig1}.

\begin{figure}
 \includegraphics[scale=.6,angle=-90]{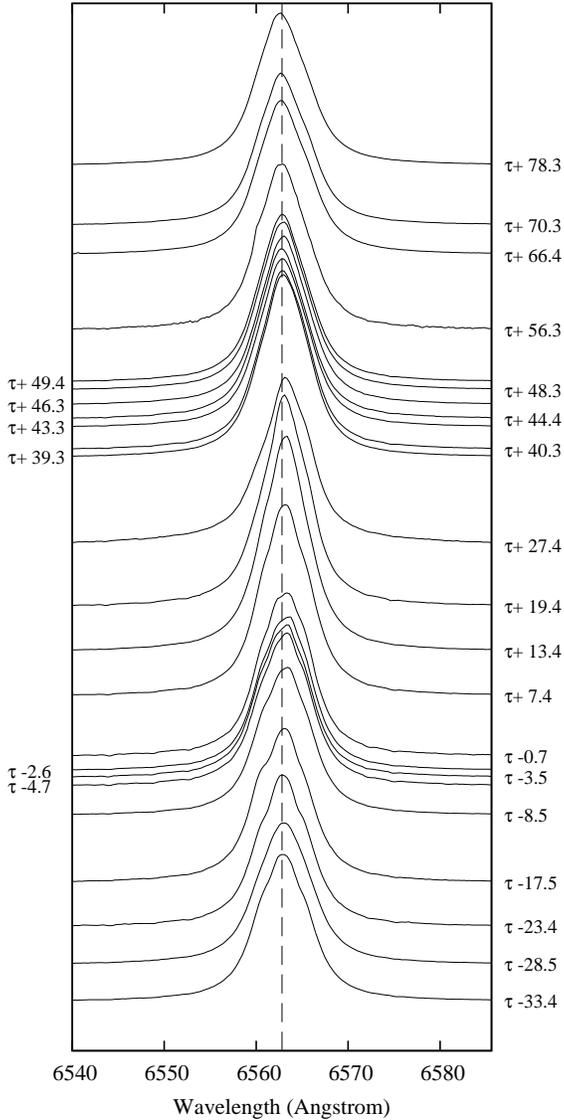}
  \caption{H$\alpha$ emission line over the period of observation (excluding 2014 May 31 and June 6). The vertical dashed line indicates the rest wavelength. The observations are ordered from earliest to latest from bottom to top. Heliocentric corrections have been applied. Here $\tau$ is the time of periastron.}
 \label{fig:fig2}
\end{figure}

Fig.~\ref{fig:fig2} shows the H$\alpha$ emission line during the observation period, ordered earliest to latest from bottom to top (excluding 2014 May 31 and June 6 due to the instrumental problems). The \halpha\ line remains single peaked during all observations, though the line does exhibit a complicated structure, and is often asymmetric, most often exhibiting a stronger blue component.  This asymmetry is typically observed in binary Be stars and is interpreted as a blending of the double peaked line originating from the circumstellar disc \citep[e.g.][]{hanuschik88}. 

\begin{figure}
 \includegraphics[scale=.6,angle=-90]{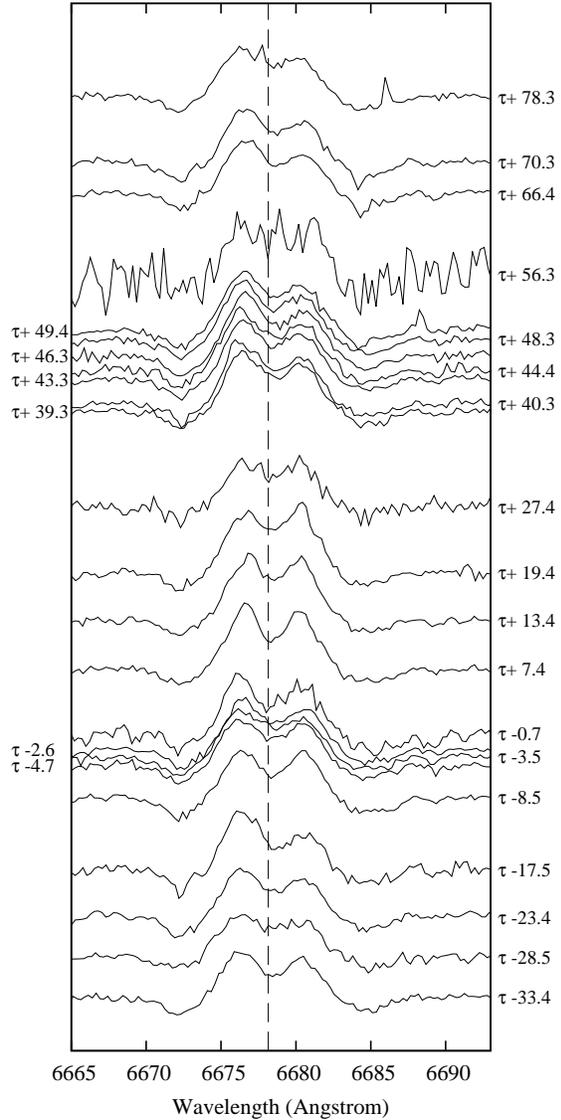}
  \caption{He~I emission line over the period of observations (excluding 2014 May 31 and June 6). The vertical dashed line indicates the rest wavelength. The observations are ordered from earliest to latest from bottom to top. Heliocentric corrections have been applied. The wider He~I absorption is just apparent around the double emission feature.}
 \label{fig:fig3}
\end{figure}

Similarly, Fig.~\ref{fig:fig3} shows the He~I line, which shows a consistent double peak structure for all observations.  Here too, the line structure is asymmetric, with the majority of observations demonstrating a stronger blue component (see discussion below). Due to the lower signal to noise ratio the He~I line is very poorly resolved for observations on 2014 May 31, June 6 and 29 (27.4, 33.3 and 56.3 d after periastron, respectively). The double peaked line associated with the circumstellar disc lies within a broader absorption feature associated with the star.

The equivalent width\footnote{By definition the equivalent width of an emission line is always negative. In this paper, we will refer to the absolute value of the equivalent width, and therefore a larger (smaller) value of $|W_\lambda|$ denotes a stronger (weaker) emission line.} of the \halpha\ line has been measured by integrating over the line with the standard IRAF procedures. Since the position of the continuum must be estimated from the data, in order to estimate the error in the continuum selection the measurement was made more than once and the average calculated value is stated. It is assumed that the H$\alpha$ absorption is negligible.

The equivalent width of the He~I line was measured by integrating over the line using the standard IRAF procedures. Since the He~I emission lines are visible in all observations, the exact shape of the underlying absorption feature is unknown. The equivalent width was therefore measured from the base of the He~I emission line, where it deviates from the absorption feature. Given the lower signal-to-noise ratio of the He~I line, the uncertainty in the base position is higher. Again the measurement was performed more than once and the average result is stated.

Further, to measure the variation in the double peak structure and the peak separation of the He~I line, the Violet (V) and Red (R) components were fitted using two Gaussian lines, relative to the continuum.\footnote{The line fitting was performed using the {\sc fityk} program \citep{wojdyr10}.} The resulting V/R variation, the ratio of the height of the V component to R component (relative to the continuum), and the change in peak separation is determined from these Gaussian fits.  The fits are also used as a second measure of the equivalent width of the He~I line, though the resulting absolute value is slightly lower than that found through integration from the base. This provides a better comparison to the results reported in \citet{chernyakova14}. 

The statistical errors in the equivalent widths were estimated using \citep{vollmann06},
\[
 \sigma\left(W_\lambda\right) = \sqrt{1+\frac{\bar{F}_c}{\bar{F}}} \frac{\Delta \lambda - W_\lambda}{S/N},
\]
where $\bar{F}_c$ is the average flux of the continuum, $\bar{F}$ is the average flux of the line, $\Delta \lambda$ is the width of the emission line, and $S/N$ is the signal to noise ratio. For the line integration measurements the line continuum average ($\bar{F}_c$) was taken from the fitted background at the central wavelength and the average flux ($\bar{F}$) was measured in IRAF for each observation.  An average value for $\Delta \lambda$ was estimated from the measurements and used for all error calculations, with the exception of 2014 May 31 and 2014 June 6 where larger values for $\Delta \lambda$ were used due to the broader line profile introduced by the instrumental problems. The standard deviation of the different measurements performed to obtain the average value are also included in the reported error.  However, this error was generally smaller than the statistical error. For the Gaussian fitted lines, the parameters were taken from the continuum and line fits.

\psrbl\ was also observed on 2014 May 21 and 22 with the SAAO~1.9-m telescope (three times per night), using the grating spectrograph. The spectrograph was used in a broad band configuration, with a wavelength coverage of $\sim 3100 - 7600$~\AA\ and a resolution of $\sim5$~\AA. The results are shown in Figure~\ref{fig:fig4} as the two grey triangles, where the uncertainty is due only to the standard deviation of the three measurements.

\section{Results}

Fig.~\ref{fig:fig4} shows the resulting light curves for the equivalent widths of the H$\alpha$ and He~I lines, and the associated properties. The results are tabulated in Table~\ref{tab:SALT}. The vertical solid line shows the time of periastron, while the two dashed vertical lines correspond to the last and first detection of the pulsar  around the 2010 periastron passage. 

\subsection{H$\alpha$ line}

\begin{figure}
 \includegraphics[scale=.63]{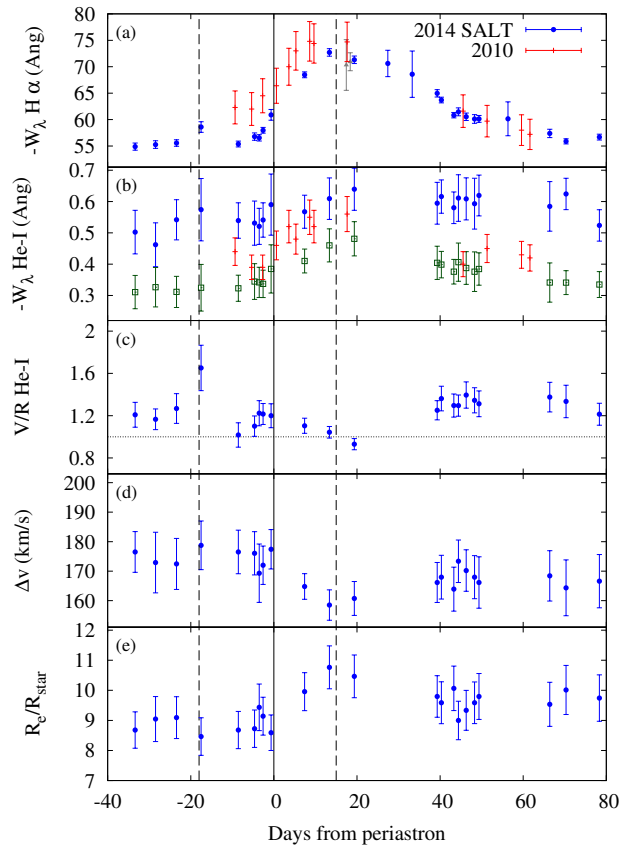}
\caption{Results obtained from the RSS spectroscopic observations around the 2014 periastron passage (blue circles) compared to the observations around the 2010 periastron passage \citep{chernyakova14} (red crosses). The two dashed vertical lines correspond to the last and first detection of the pulsar around the 2010 periastron passage, while the vertical solid line indicates periastron. (a) Equivalent width of the H$\alpha$ line. The grey triangles indicate the observations made with the SAAO 1.9-m. (b) The equivalent width of He~I measured from the base of the emission line (solid blue circles) and from the line profile fit to the continuum (open green squares). (c). The V/R in the He~I line. (d) The peak separation of the He~I line.  (e) The emission location of He~I assuming a Keplerian disc (equation~\ref{eqn:kep_disc}).}\label{fig:fig4}
\end{figure}

Fig.~\ref{fig:fig4}-(a) shows the equivalent width of the H$\alpha$ line from the period around the 2014 periastron obtained with SALT (blue circles) compared to the measurements around the 2010 passage \citep[red crosses,][]{chernyakova14}. The grey triangles show the measurements obtained with the SAAO 1.9-m. 
The average of the H$\alpha$ equivalent width before the first disc plane crossing (earlier than -20~d) is  $\langle W_{{\rm H}\alpha}\rangle=-55.2\pm0.8$~\AA. This is consistent with what has been previously reported, for example, $-54\pm2$~\AA\ \citep{negueruela11},  $-50\pm5$~\AA\ and $-49\pm2$~\AA\ \citep{vansoelen12} approximately $-474$, $+116$ and $+507$~d from the 2010 periastron passage.  

The orbital variation in the equivalent width follows the same trend as previously observed, with an increase in the absolute equivalent width near to periastron, increasing towards and peaking after periastron. The maximum $|W_\lambda|$ occurs $\sim 13$~d after periastron, and appears to be decreasing by $\sim 19$~d after periastron. After 40~d from periastron there is a smooth decrease down to the pre-periastron level. 

There is also an increase in the equivalent width near the expected time of the first disc plane crossing at $\sim -17$~d from periastron. The observation shows a distinct increase in the asymmetry of the H$\alpha$ line, with an increase in the blue component of the line.

The equivalent width measurements before periastron are lower than those reported during 2010/2011, but are comparable around the peak and after periastron. The observation immediately following the onset of the 2014 \textit{Fermi} flare ($\tau=+33$~d, 2014 June 6) suffered from degradation due to instrumental difficulties resulting in an increased uncertainty in the continuum position and also an effective loss of light. The larger error shown in Fig.~\ref{fig:fig4}-(a) is due to the lower signal to noise, but the actual uncertainty may be higher given that an unmeasurable component of light is lost during this period. No conclusion can or should be drawn from this point, which is presented only for the sake of completeness.

\subsection{He~I line}

\subsubsection{Equivalent width}

The equivalent widths of the He-I line around the 2014 periastron are shown in Fig.~\ref{fig:fig4}-(b).  The closed blue circles denote the equivalent width as measured by integrating from the base of the line, while the open green squares denote the measurement obtained from the Gaussian fit, relative to the continuum.  Shown for comparison are the measurements around 2010 reported by \citet{chernyakova14} (red crosses). As with the H$\alpha$ line, the He~I line shows a general increase in the absolute equivalent width which peaks after periastron, though the growth is not as great. The variation in the strength of the line is consistent with the variability observed in the \halpha\ line.

\subsubsection{V/R}

In general, the violet component of the double peaked He~I line is more dominant, as is seen in the equivalent width $|W_\lambda|$ of the violet component and the ratio of peaks of the violet to red (V/R) components.  The V/R variation (Fig.~\ref{fig:fig4}-(c)) has been measured from the ratio of the heights of the Gaussian fits to the Violet and Red component of the He~I line. The height is measured relative to the continuum and the error is given by the statistical error of the fit. Around the first disc plane crossing, there is a marked increase in the asymmetry of the line, while there is a decrease in V/R around periastron and a small increase in the red component around the second disc plane crossing. This clearly illustrates the asymmetry of the He~I line and its variation around periastron. The V/R variation lies between $0.93<\rmn{V/R}<1.65$ during the observed period.

\section{Discussion}

Around periastron the equivalent width smoothly increased to a maximum of $-72.7\pm0.7$ around $\sim\tau+13$~d, after which it decreased to values consistent with the pre-periastron levels.  The equivalent widths after $\sim\tau+13$~d are consistent with the previous periastron passage \citep{chernyakova14} but appear to be consistently lower before this.  This is clearly seen before periastron where the average equivalent width (and error) of the H$\alpha$ line was $-62.9\pm 3.1$~\AA{} in 2010, but $-56.7\pm0.6$~\AA{} in 2014. We interpret this as being due to variability in the Be star circumstellar disc which could well change from periastron to periastron.  This may imply that the mass of the circumstellar disc was lower before the 2014 periastron passage (as compared to the 2010 periastron passage). Since the circumstellar disc influences the shock front around the pulsar, the differences in the circumstellar disc properties will be an important contributing factor to the variations in the multi-wavelength emission observed around different periastra.   

There is also a marked change in the disc around the time of the first disc plane crossing ($\tau-17.5$~d), as is noted by the increase in the strength of the H$\alpha$ line ($|\Delta W_{\rmn{H}\alpha}| = 3.0\pm 1.2$~\AA), the increase in the asymmetry of the H$\alpha$ line (see Fig.~\ref{fig:fig2}) and the change in the V/R variation of the He~I line. The disruption of the circumstellar disc around periastron is expected due to tidal interaction as well the collision between the disc and the pulsar wind \citep{okazaki11} and these observations around the disc plane crossing, point to rapid changes in the disc structure around this period. 

The disc is believed to be truncated around periastron and, therefore, the interaction around the second  disc plane crossing should not introduce as dramatic a change in the observed profile, and no sudden change in the H$\alpha$ or He~I lines are observed around this orbital phase.  However, it should be noted that the observations were taken $\sim2$~days after the expected time of the end of the pulsar eclipse.  

The peak separation, $\Delta v$, of the He~I line follows a smooth evolution around periastron as shown in Fig~\ref{fig:fig4}-(d). The separation has been calculated from the difference between the centres of the Gaussian fits to the V and R components of the He~I line, assuming a central wavelength of 6658.15~\AA.  The average separation is $170$~km~s$^{-1}$ over the observed period, varying between $158<\Delta v< 181$~km~s$^{-1}$, with the smallest separation occurring around the peak in the equivalent widths. 

For a Keplerian circumstellar disc, the decrease in $\Delta v$ implies the dominating region of emission occurs further out in the circumstellar disc as the emission line strength increases. From the measured peak separation we have placed a constraint on the location of the emitting region, which has been determined using \citep{huang72}
\begin{equation}
 \frac{R}{R_\star} = \left( \frac{2 v \sin i}{\Delta v} \right)^2,
\label{eqn:kep_disc}
\end{equation}
where $R_\star$ is the radius of the Be star. Using $v \sin i = 260\pm15$~km~s$^{-1}$ \citep{negueruela11} we find that the location of the emission region varies between $8.3 < R/R_\star < 10.8$ around the periastron period (Fig.~\ref{fig:fig4}-(e)).

\section{Conclusion}

RSS spectra obtained with SALT around the 2014 periastron passage of \psrbl\ have shown  similar results to what was previously observed from the system.  Both the H$\alpha$ and He~I lines show variability around this period, with $|W_{\lambda}|$ increasing to a maximum slightly after periastron, and a general decrease following this.  The average $|W_{\lambda}|$ before periastron is lower than what was previously observed, which we attribute to variability of the Be star's circumstellar disc between periastron passages. There is an indication of a change in the H$\alpha$ equivalent width and the He~I V/R variation around the period of the first disc plane crossing. And while no data is obtained immediately before the reported onset of the \textit{Fermi} flare, the time following this shows a smooth decay during a period of ongoing GeV gamma-ray emission.  Additional multi-wavelength observations can place further constraints on the processes occurring in this system around periastron \citep[see e.g.][]{chernyakova15}.

\section*{Acknowledgements}

Observations reported in this paper were obtained with the Southern African Large Telescope (SALT) under programs 2013-2-RSA-012 and 2014-1-RSA-001. This paper uses observations made at the South African Astronomical Observatory (SAAO). PV acknowledges support from the NRF of South Africa.

\begin{table*}
\begin{minipage}{160mm}
\centering
\caption{Measurements obtained with SALT. } 
\label{tab:SALT}
\begin{tabular}{lccccc}\hline
MJD & day from periastron &  $W_{\rmn{H}\alpha}$ & $W_{\rmn{He~I}}$  & V/R  & $\Delta v$  \\
 \\   & d & \AA & \AA & \AA & km/s \\ \hline      
56748.0 & -33.4 & $ -54.9 \pm 0.7 $ & $ -0.50 \pm 0.07 $ & $ 1.21 \pm 0.12 $ & $ 176.5 \pm 6.9 $ \\
56752.9 & -28.5 & $ -55.3 \pm 0.7 $ & $ -0.46 \pm 0.07 $ & $ 1.17 \pm 0.10 $ & $ 172.9 \pm 10.3 $ \\
56758.0 & -23.4 & $ -55.6 \pm 0.6 $ & $ -0.54 \pm 0.06 $ & $ 1.27 \pm 0.14 $ & $ 172.5 \pm 8.6 $ \\
56763.9 & -17.5 & $ -58.6 \pm 1.0 $ & $ -0.57 \pm 0.10 $ & $ 1.65 \pm 0.21 $ & $ 178.8 \pm 8.2 $ \\
56772.9 & -8.5 & $ -55.4 \pm 0.6 $ & $ -0.54 \pm 0.06 $ & $ 1.02 \pm 0.12 $ & $ 176.5 \pm 7.4 $ \\
56776.8 & -4.7 & $ -56.8 \pm 0.7 $ & $ -0.53 \pm 0.07 $ & $ 1.10 \pm 0.10 $ & $ 176.1 \pm 7.3 $ \\
56777.9 & -3.5 & $ -56.5 \pm 0.6 $ & $ -0.52 \pm 0.06 $ & $ 1.22 \pm 0.12 $ & $ 169.3 \pm 9.9 $ \\
56778.8 & -2.6 & $ -58.0 \pm 0.5 $ & $ -0.54 \pm 0.05 $ & $ 1.22 \pm 0.10 $ & $ 172.0 \pm 6.5 $ \\
56780.8 & -0.7 & $ -60.9 \pm 1.0 $ & $ -0.59 \pm 0.10 $ & $ 1.20 \pm 0.11 $ & $ 177.4 \pm 6.7 $ \\
56788.8 & 7.4 & $ -68.5 \pm 0.5 $ & $ -0.57 \pm 0.05 $ & $ 1.10 \pm 0.07 $ & $ 164.8 \pm 4.4 $ \\
56794.8 & 13.4 & $ -72.7 \pm 0.7 $ & $ -0.61 \pm 0.07 $ & $ 1.04 \pm 0.05 $ & $ 158.5 \pm 5.2 $ \\
56800.8 & 19.4 & $ -71.3 \pm 0.7 $ & $ -0.64 \pm 0.07 $ & $ 0.93 \pm 0.05 $ & $ 160.7 \pm 5.7 $ \\
56808.8 & 27.4 & $ -70.6 \pm 2.5 $ & \\ 
56814.7 & 33.3 & $ -68.6 \pm 4.4 $ &            \\
56820.7 & 39.3 & $ -65.0 \pm 0.7 $ & $ -0.59 \pm 0.07 $ & $ 1.25 \pm 0.09 $ & $ 166.1 \pm 6.8 $ \\
56821.7 & 40.3 & $ -63.7 \pm 0.5 $ & $ -0.62 \pm 0.05 $ & $ 1.36 \pm 0.12 $ & $ 167.9 \pm 7.4 $ \\
56824.7 & 43.3 & $ -60.8 \pm 0.5 $ & $ -0.58 \pm 0.05 $ & $ 1.30 \pm 0.11 $ & $ 163.9 \pm 7.5 $ \\
56825.8 & 44.4 & $ -61.5 \pm 0.8 $ & $ -0.61 \pm 0.07 $ & $ 1.30 \pm 0.10 $ & $ 173.4 \pm 7.2 $ \\
56827.7 & 46.3 & $ -60.5 \pm 0.7 $ & $ -0.61 \pm 0.07 $ & $ 1.40 \pm 0.12 $ & $ 170.2 \pm 7.0 $ \\
56829.7 & 48.3 & $ -60.1 \pm 0.8 $ & $ -0.59 \pm 0.08 $ & $ 1.35 \pm 0.12 $ & $ 167.9 \pm 7.3 $ \\
56830.8 & 49.4 & $ -60.1 \pm 0.6 $ & $ -0.62 \pm 0.07 $ & $ 1.31 \pm 0.12 $ & $ 166.1 \pm 8.7 $ \\
56837.8 & 56.3 & $ -60.1 \pm 3.2 $ & \\
56847.8 & 66.4 & $ -57.4 \pm 0.8 $ & $ -0.58 \pm 0.08 $ & $ 1.38 \pm 0.14 $ & $ 168.4 \pm 8.5 $ \\
56851.7 & 70.3 & $ -55.9 \pm 0.5 $ & $ -0.62 \pm 0.05 $ & $ 1.33 \pm 0.15 $ & $ 164.3 \pm 9.5 $ \\
56859.8 & 78.3 & $ -56.7 \pm 0.6 $ & $ -0.52 \pm 0.05 $ & $ 1.21 \pm 0.10 $ & $ 166.6 \pm 9.0 $ \\
 \hline
\end{tabular}
\end{minipage}

\end{table*}

\bsp

\label{lastpage}

\end{document}